\documentclass[aps,prl,twocolumn,showpacs,groupaddress]{revtex4}
\usepackage{amssymb}
\usepackage{graphicx}
\usepackage{dcolumn}
\usepackage{bm}
\usepackage{amsmath}

\begin{document}

\title{Coherent control with broadband squeezed vacuum}
\author{Barak Dayan}
\email[]{Barak.Dayan@Weizmann.ac.il}
\author{Avi Pe'er}
\email[]{Avi.Peer@Weizmann.ac.il}
\author{Asher A. Friesem}
\author{Yaron Silberberg} \homepage[]{www.weizmann.ac.il/home/feyaron/}
\affiliation{Department of Physics of Complex Systems, Weizmann Institute of Science,\\
Rehovot 76100, Israel}


\begin{abstract}
We report the experimental demonstration of coherent control with
high power, broadband squeezed vacuum. Although incoherent and
exhibiting the statistics of a thermal noise, broadband squeezed
vacuum is shown to induce certain two-photon interactions as a
coherent ultrashort pulse with the same spectral bandwidth.
Utilizing pulse-shaping techniques we coherently control the
sum-frequency generation of broadband squeezed vacuum over a range
of two orders of magnitude. Coherent control of two-photon
interactions with broadband squeezed vacuum can potentially obtain
spectral resolutions and extinction ratios that are practically
unattainable with coherent pulses.
\end{abstract}


\maketitle

The quantum mechanical expressions for broadband two-photon
interactions clearly reflect the underlying spectral quantum
interference. For example, the final population  of an atomic
level with energy $\Omega$, due to a two-photon absorption (TPA)
induced by light with spectral amplitude $E(\omega)$ is
\cite{PRA60}:

\begin{eqnarray} \label{Ew}
p_f \propto \bigg | \int E(\Omega/2+\xi)E(\Omega/2-\xi)\: d\xi \
\bigg| ^{\:2}
\end{eqnarray}

This expression exhibits quite directly the fact that TPA may
occur by photon pairs with energies $\Omega/2+\xi$ and
$\Omega/2-\xi$, for any $\xi$ within the bandwidth of the light.
For coherent light, $E(\omega)$ has a defined spectral phase:
$E(\omega)=A\left(\omega\right)\exp\left[i \Theta(\omega)\right]$.
Consequently, the final population can be controlled by applying a
spectral phase filter $\Phi(\omega)$ to the light, making the
quantum interference constructive or destructive, as desired:
\begin{eqnarray} \label{Aw}
p_f \propto \bigg | \int &&
\hspace{-0.5cm} A(\Omega/2+\xi)A(\Omega/2-\xi) \nonumber \\
&\times& \exp\left[i \Theta\left(\Omega/2+\xi\right) +
i \Theta\left(\Omega/2-\xi\right)\right] \nonumber \\
&\times& \exp[i \Phi(\Omega/2+\xi) + i \Phi(\Omega/2-\xi)] \: d\xi
\ \bigg| ^{\:2}
\end{eqnarray}

Typically, pulses with a constant spectral phase
(transform-limited pulses) maximize nonresonant interactions, as
becomes evident if we let $\Theta(\omega)=\Phi(\omega)=0$ in Eq.
(\ref{Aw}). The technique of coherent control
\cite{Tannor&Rice,Shapiro&Brumer,Warren&Rabitz,Rice&Zhao} usually
exploits pulse-shaping techniques \cite{Weiner_Shaping} to apply a
spectral phase filter to mode-locked ultrashort pulses, in order
to manipulate the quantum interference and steer multiphoton
interactions towards desired states. In particular, it was
demonstrated \cite{Nature396} that applying a spectral phase
filter that is anti-symmetric about $\Omega/2$ does not affect the
TPA probability, although it may significantly stretch the pulse
and lower its peak power. The simple reason for this effect is
that when $\Phi(\Omega/2+\xi)=- \Phi(\Omega/2-\xi)$, opposite
phases are applied to the complementary modes in all the
mode-pairs that contribute to the TPA process. Thus, as is evident
from Eq. (\ref{Aw}), the overall phase contribution of the phase
filter is cancelled out, leaving the efficiency equal to that of a
transform-limited pulse. Nonetheless, a symmetric phase filter
does effect the TPA probability, and can even eliminate it.
Similar coherent control was demonstrated over sum-frequency
generation (SFG) with ultrashort pulses, establishing that SFG in
thick nonlinear crystals may be considered equivalent (in the
perturbative limit) to TPA \cite{Weiner&Zheng1,Weiner&Zheng2}.\\

Here we establish that coherent control can be performed with
incoherent, broadband squeezed vacuum, generated by a parametric
down-conversion \cite{Mandel&Wolf,Hong&Mandel} of a narrowband
pump laser in a nonlinear crystal. We show that squeezed vacuum
with a spectral bandwidth that exceeds a certain limit induces TPA
and SFG just like a coherent ultrashort pulse with the same
spectral bandwidth. This effect occurs as long as the final state
energy lies within the spectrum of the pump laser that generated
the squeezed vacuum. Consequently, such interactions can be
coherently controlled by pulse-shaping techniques, even though the
squeezed vacuum is neither coherent nor pulsed, but rather is
incoherent and exhibits the properties of a broadband thermal
noise \cite{Mandel&Wolf,Scully&Zubairy}. We prove this principle
experimentally by coherently controlling the SFG signal induced by
high-power broadband squeezed vacuum, and obtaining similar
results to those obtained with coherent ultrashort pulses.\

The underlying principle of our coherent control scheme is that
while coherence of all the spectral paths is indeed required for
coherent control, it does not necessarily imply coherence of the
inducing light. Since the quantum interference that governs
two-photon interactions always involves pairs of photons, it
requires the photon-\textsl{pairs} to be coherent. In other words,
as is obvious from Eq. (\ref{Ew}), it does not matter whether
$E(\omega)$ has a defined phase for every $\omega$, but rather
whether the product $E(\Omega/2+\xi)E(\Omega/2-\xi)$ has a defined
phase for every $\xi$. Although broadband squeezed vacuum has a
random spectral phase, it does exhibit exactly this phase behavior
at frequency-pairs due to the inherent quantum correlations within
its spectrum. The two-mode squeezing which occurs during the
down-conversion process leads to complete amplitude correlations
and phase anti-correlations between the spectral-components of
$E(\omega)$ at complementary frequency-pairs that sum to the pump
frequency \cite{Mollow&Glauber,McNeil&Gardiner}:

\begin{eqnarray} \label{correlations}
{ \textstyle \frac{1}{\sqrt{ \omega_p/2+\xi}} }\:
A(\omega_p/2+\xi) & \sim &
\textstyle{\frac{1}{\sqrt{\omega_p/2-\xi}}} \: A(\omega_p/2-\xi) \nonumber \\
\Theta (\omega_p/2+\xi ) &\sim& \pi/2 - \Theta(\omega_p/2-\xi)
\end{eqnarray}

These correlations drastically affect the quantum interference
that governs two-photon interactions. Generally, the expected
efficiency of a broadband thermal noise at inducing nonlinear
interactions is extremely low, and usually ultrashort pulses with
high peak-powers must be used. However, for TPA (or SFG) with
down-converted light, the situation is strikingly different when
the final state energy is equal to the pump frequency. As becomes
clear if one combines equations \ref{Aw} and \ref{correlations}
taking $\Omega=\omega_p$, the random phase of $E(\Omega/2+\xi)$ is
always compensated by the opposite phase of $E(\Omega/2-\xi)$.
Thus, the overall phase contribution of every mode pair is
cancelled out, and the integrand of Eq. (\ref{Aw}) is affected
only by the phase-filter $\Phi(\omega)$, exactly as if the
interaction was induced by a transform-limited pulse with the same
spectrum:

\begin{eqnarray} \label{TL}
p_f \propto \bigg | \int&& \hspace{-0.5cm}
A(\Omega/2+\xi)A(\Omega/2-\xi) \nonumber \\
&\times& \exp[\Phi(\Omega/2+\xi) + \Phi(\Omega/2-\xi)] \: d\xi \
\bigg| ^{\:2}\:.
\end{eqnarray}

Naturally, the cancelling out of the phase of the squeezed light
resembles the cancelling out of an anti-symmetric phase
manipulation on a coherent ultrashort pulse. This similarity may
be further clarified by considering cw-pumped broadband squeezed
vacuum as similar to a classical ultrashort pulse, which has
undergone a spectral phase manipulation that is random at every
mode, but antisymmetric about $\frac{\omega_p}{2}$. Such a
hi-resolution phase manipulation, which is practically
unattainable by pulse-shaping techniques, will stretch the pulse
to a continuous wave. Consequently, the pulse's ability to induce
TPA or SFG will be reduced drastically, except for final states
with energy that is exactly $\omega_p$, where the full efficiency
of the process will remain unaffected. This constructive
interference occurs only when the final state energy falls within
the spectrum of the pump laser. As was indeed observed by Abram
{\sl et al.} \cite{Abram&Raj}, this implies that the spectrum of
the SFG signal actually reproduces the spectrum of the pump laser.
Note that in the case of a single-frequency pump laser, this
implies a possible spectral resolution of a few KHz. Such a
spectral resolution is phenomenally high for an interaction that
is induced by light with spectral bandwidth that is typically many
orders of magnitude wider.\\

A full analytic quantum mechanical calculation \cite{Calculations}
reveals a somewhat more complicated behavior. The SFG or TPA
signal at the pump frequency is composed of two parts, which may
be referred to as the 'quantum term' and the 'classical term':
$I_\Omega=I_\Omega^{c}+I_\Omega^{q}$. While the quantum term
results from the coherent summation of conjugated spectral
components, the classical term results from the incoherent
summation of all random spectral combinations. The quantum term is
therefore the one that is equivalent to a coherent pulse and can
be coherently controlled. The classical term, however, is a direct
result of the incoherence of the down-converted light and
therefore it is unaffected by spectral-phase manipulations. Thus,
it may be regarded as an incoherent background noise, which limits
the equivalence of the down converted light to a coherent pulse.\

The ratio between the quantum term and the classical term can be
approximated by:
\begin{eqnarray} \label{ratio}
\frac{I_\Omega^{q}}{I_\Omega^{c}} \approx
\frac{B}{2(\gamma_p+\gamma_f)}\frac{n^2+n}{n^2} \: ,
\end{eqnarray}
where $n$ is the spectral average of the mean photon flux, and
$B,\gamma_p,\gamma_f$ are the bandwidths of the squeezed vacuum,
the pump laser and the final state respectively (in the case of
SFG $\gamma_f$ represents the spectral resolution of the
measurement). The factor of $2$ results from the assumed collinear
configuration. This expression reveals the importance of using
spectrally broad down-converted light. The quantum term becomes
dominant only when the down-converted bandwidth exceeds both the
pump bandwidth and the spectral resolution of the measurement:
$B>2(\gamma_p+\gamma_f)(\frac{n^2}{n^2+n})$. Moreover, Eq
(\ref{ratio}) shows that the quantum term exhibits a linear
intensity dependence at low powers, as was indeed observed by
Georgiades {\it et al.} \cite{Georgiades&Kimble}.\\

\begin{figure}[tbp] \label{fig1}
\begin{center}
\includegraphics[width=8.6cm]{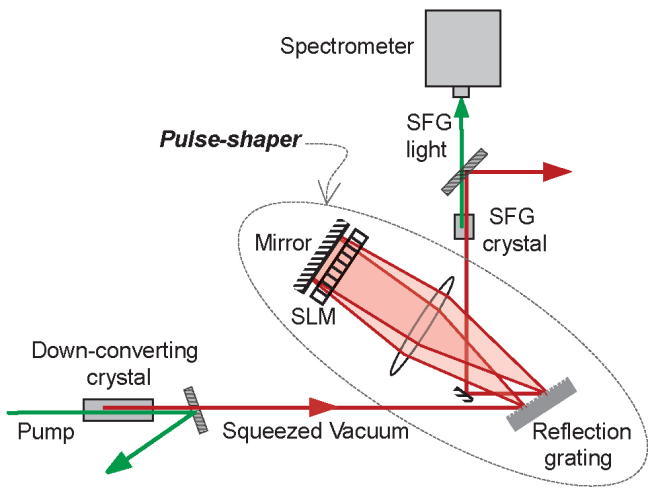}
\end{center}
\caption{Experimental system for coherent control of SFG with
broadband squeezed vacuum. Down-converted light with spectral
bandwidth of ~60 nm around 1064 nm is generated in a 9 mm long
periodically-poled KTP nonlinear crystal pumped by 1 mJ, 8-ns
pulses at 532 nm from a doubled Q-switched Nd:YAG laser. The
remainder of the pump beam is removed after the crystal by a
filter. The squeezed vacuum then undergoes spectral phase
manipulations in a folded pulse-shaper that is composed of a
reflection grating, a lens, a 128-element liquid-crystal spatial
light-modulator (SLM) and a mirror. In this configuration, a
single set of reflection grating and lens is used both to image
the different spectral components of the incoming beam on the
phase elements of the SLM, and to reassemble those components when
they are reflected back by the mirror behind the SLM. The output
beam is directed to a second, 2 mm long, periodically-poled KTP
crystal, where the SFG process occurs. The SFG signal is then
measured by a spectrometer with 0.03 nm spectral resolution.}
\end{figure}

To experimentally demonstrate these principles we used a
programmable pulse-shaper to apply a spectral phase filter to
broadband (60 nm centered at 1064 nm) down-converted light,
emitted from a nonlinear PPKTP crystal pumped by spectrally narrow
($\approx$0.01nm) 8-ns pulses at 532nm (Fig. \ref{fig1}). The
light was directed from the pulse-shaper to a second PPKTP
crystal, and the resulting SFG signal was measured by a
spectrometer with a spectral resolution of 0.03 nm. Our
calculations show that the behavior of the SFG signal in this
scheme is equivalent to TPA with final level broadening equal to
the spectral resolution of the spectrometer. \\

Figures 2 and 3 show the experimental and the calculated results
of our coherent control experiments with high-power broadband
squeezed vacuum. Figure 2(a) depicts the measured (circles) and
calculated (line) spectrum of the SFG signal on a logarithmic
scale, obtained without any spectral phase manipulation, save for
dispersion compensation. In this measurement the pulse shaper was
set only to compensate for the overall dispersion of the system.
The graph clearly shows the remarkable difference between the
quantum peak at 532nm, which is as narrow as our spectral
resolution, and the broad, two orders of magnitude lower,
classical background.\\

\begin{figure}[tbp]\label{fig2}
\begin{center}
\includegraphics[width=8.6cm]{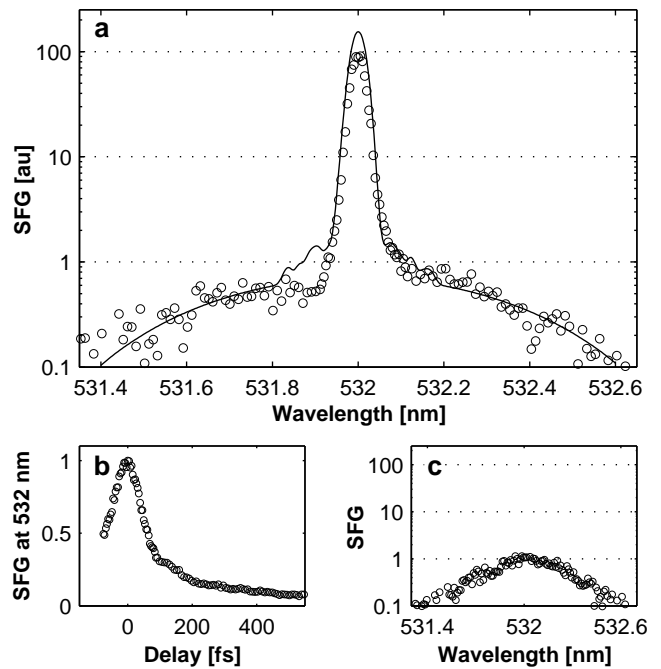} \end{center}
\caption{Coherent control of SFG with broadband squeezed vacuum by
a linear spectral phase filter that is equivalent to a delay
between the higher and lower halves of the spectrum. (a)
Experimental (circles) and calculated (line) SFG spectrum without
any delay. In this measurement the pulse shaper was set only to
compensate for the dispersion of the optical system. (b) The
measured SFG signal at 532 nm as a function of the equivalent
delay between the two spectral halves of the down-converted light.
(c) The experimental SFG spectrum at an equivalent delay of 1.5
ps, where the quantum term is completely suppressed.}
\end{figure}

To demonstrate coherent control of the SFG process we applied a
linear phase function to the lower half of the spectrum, and a
linear phase function with opposite slope to the higher half of
the spectrum: $\Phi(\omega)=\tau \, | \, \omega-\omega_p/2 \, |$.
Such a phase function is equivalent to delaying one half of the
spectrum by $2 \, \tau$ relative to the other. Figure 2(b) shows
the measured SFG signal at 532 nm as a function of the overall
delay in fs. The rapid decay in the signal verifies the predicted
equivalence of the down-converted light to an ultrashort pulse of
$\sim$ 65 fs, despite the fact that the actual duration of the
down-converted pulse is 5 orders of magnitude longer. Figure 2(c)
shows the SFG spectrum at an equivalent delay of 1.5 ps, where the
quantum term is completely suppressed, leaving only the unaffected
classical background.\\

Next, we applied a sinusoidal phase filter $\Phi(\omega)=\alpha \,
\sin\left( \beta \left(\omega-\frac{1}{2}\omega_p\right)+\theta
\right)$ to the down-converted spectrum. When $\theta=0,\pm \pi,
\pm 2 \pi, \, ...$ this function is anti-symmetric about
$\frac{1}{2}\omega_p$ and therefore, as noted earlier, it does not
affect the SFG signal at $\omega_p$. On the other hand, for
$\theta=\pm \frac{1}{2}\pi, \pm \frac{3}{2} \pi, \, ...$ the
filter is a symmetric function about $\frac{1}{2}\omega_p$, and
with the appropriate values of $\alpha$ and $\beta$ it is expected
to completely suppress the quantum part of the SFG signal. Figure
3 depicts the measured (circles) and the calculated (line) SFG
signal at 532 nm as a function of the phase $\theta$, showing the
expected periodic reconstruction of the full signal at
$\theta=0,\pm \pi, \pm 2 \pi \,$, and a suppression of the signal
at $\theta=\pm \frac{1}{2}\pi, \pm \frac{3}{2} \pi \,$. The SFG
signal is reduced then to about $13.5\%$ of the maximal value.
This should be compared with the calculated value of $\sim 1\%$,
being the classical background level. We believe the residual
signal at the minima points is due to large shot-to-shot
fluctuations in the down-converted spectral envelope, which
affected the averaged measurement of the SFG
signal.\\

\begin{figure}[tbp]
\begin{center}
\includegraphics[width=8.6cm]{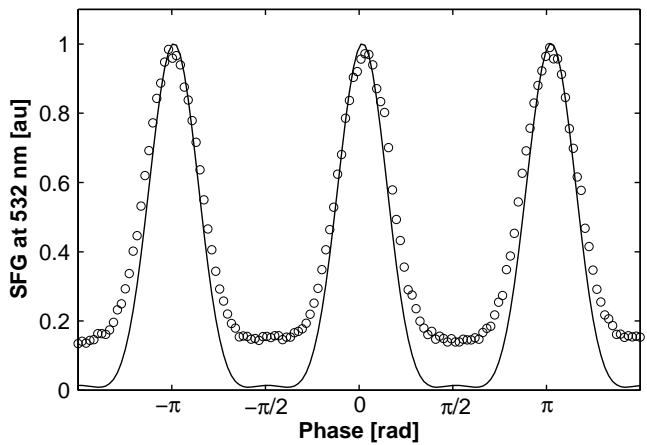}\end{center}
\caption{Coherent control of SFG with broadband squeezed vacuum by
a periodic spectral phase function $\Phi(\omega)=\alpha \,
\sin\left( \beta
\left(\omega-\frac{1}{2}\omega_p\right)+\theta\right)$.
Experimental (circles) and calculated (line) SFG signal at 532 nm
as a function of the phase $\theta$. The signal remains unaffected
when $\Phi(\omega)$ is anti-symmetric about $\omega_p/2$, and
suppressed elsewhere. The values of $\alpha$ and $\beta$ were
chosen to maximize the suppression.} \label{fig3}
\end{figure}

These results verify that SFG at the frequencies of the pump laser
is induced coherently by broadband squeezed vacuum, and therefore
can be coherently controlled by pulse-shaping techniques, despite
the fact that the squeezed vacuum may be neither coherent, nor
pulsed. Accordingly, the SFG signal at the these frequencies is as
coherent as the pump laser that generated the squeezed vacuum. We
note that a similar principle is known to hold at low-power
squeezed vacuum, where the nonlocal second-order coherence effects
of entangled photon-pairs are determined by the first-order
coherence of the pump laser \cite{Ou&Mandel}. However, the
equivalence of high-power squeezed vacuum to a coherent ultrashort
pulse is not directly connected to second-order coherence, which
exhibits a similar temporal behavior only at the single-photons
regime \cite{HOM}. While non-classical features are expected in
two-photon interactions with squeezed vacuum, they were not
demonstrated in these experiments.\\

Two-photon interactions induced by broadband squeezed vacuum
exhibit the low intensity and the narrow spectral resolution of
the pump laser, while exhibiting the efficiency and temporal
resolution of an ultrashort pulse with the same broad bandwidth as
the squeezed vacuum. The possibility to induce nonlinear
interactions like an ultrashort pulse, yet with spectral
resolution and peak intensities of a continuous, single-frequency
laser may offer new opportunities for various applications such as
multi-users optical communication \cite{Communication} and
multiphoton microscopy.

\end{document}